\documentclass[slac_one]{revtex4}
\usepackage{graphicx}
\usepackage{fancyhdr}
\usepackage{epsfig}
\newcommand{\gev}{\,\, \mathrm{GeV}}

\newcommand{\MW}{M_{\rm W}}
\newcommand{\MZ}{M_{\rm Z}}
\newcommand{\MH}{M_{\rm H}}
\newcommand{\mt}{m_{\rm t}}
\newcommand{\Pf}{{\rm f}}
\newcommand{\al}{\alpha}
\newcommand{\als}{\alpha_{\rm s}}
\newcommand{\SinEff}{${\sin^2\theta^{\mbox{\footnotesize lept}}_{\mbox{\footnotesize eff}}}$}
\newcommand{\sineff}{\sin^2\theta^{\mbox{\scriptsize lept}}_{\mbox{\scriptsize eff}}}

\newcommand{\sinefff}{\sin^2\theta^{f}_{\mbox{\scriptsize eff}}}

\begin{document}

\title{{\small{2005 International Linear Collider Workshop - Stanford,
U.S.A.}} 
\hfill {\rm\small FERMILAB-Conf-05-318-T\\[-.5ex] \ \hfill
 DESY 05-122\\[-.5ex] \ \hfill WUE-ITP-2005-007\\}%
\vspace{6pt}
Indirect Determination of the Higgs Mass Through Electroweak Radiative
Corrections} 

%

\author{A.~Freitas}
\affiliation{Fermi National Accelerator Laboratory, Batavia, IL 60510-500, USA}
\author{M.~Awramik}
\affiliation{DESY, Platanenallee 6, D-15738 Zeuthen, Germany;\newline
Institute of Nuclear Physics PAS, Radzikowskiego 152, PL-31342 Cracow, Poland}
\author{M.~Czakon}
\affiliation{Institut f\"ur Theoretische Physik und Astrophysik, Universit\"at
W\"urzburg, Am Hubland, D-97074 W\"urzburg, Germany;\\
Institute of Physics, University of Silesia, Uniwersytecka 4,
  PL-40007 Katowice, Poland}

\begin{abstract} 
Electroweak precision observables allow stringent tests of the
Standard Model at the quantum level and imply interesting bounds on the mass of
the Higgs boson through higher-order loop effects. Very significant constraints
come especially from the determination of the mass of the $W$ boson and from
the effective leptonic weak mixing angle. After shortly reviewing the status of
theoretical computations of the $W$ mass, the new calculation of two-loop
corrections with closed fermion loops to the effective leptonic weak mixing
angle is discussed in detail. The phenomenological implications of the new
result are analyzed including an estimate of remaining uncertainties.
\end{abstract}

\maketitle

\thispagestyle{fancy}


\section{INTRODUCTION} 

In recent years, the Standard Model of electroweak interactions has been
confirmed experimentally with outstanding success. Not only was almost all the
particle content discovered at accelerator experiments, but their properties
and interactions have been measured with high precision, in agreement with the
model prediction. The only missing piece is the Higgs boson, which is
responsible for electroweak symmetry breaking. However, even today we can
obtain meaningful constraints on the Higgs boson mass from electroweak
precision measurements. Due to the impressive accuracy of some of these
experimental results, they are sensitive to electroweak radiative corrections
at the next-to-leading (NLO) and sometimes next-to-next-to-leading (NNLO)
level, and thus depend on the impact of the Higgs boson entering in the loops.

Two of the most important quantities in this respect are the mass of the $W$
boson, $\MW$ and the sine of the leptonic effective weak mixing angle \SinEff. 
The $W$-boson mass can be inferred from the muon decay constant $G_\mu$, which is
generated through virtual W-boson exchange, so that $G_\mu \propto 1/\MW^2$.
The effective weak mixing angle, on the other hand, reflects the ratio of the
vector and axial-vector couplings, $v_f$ and $a_f$, of the $Z$ boson to fermions
($f$) at the $Z$ boson pole:
\begin{equation}
  \label{eq:def}
  \sinefff = \frac{1}{4}
  \left(1+{\rm Re} \frac{v_f}{a_f}\right).
\end{equation}
Since these couplings can be measured most
precisely for leptons, the \emph{leptonic} effective weak mixing angle \SinEff\
is usually taken as a reference. 

The current experimental world average for the $W$-boson mass is $\MW = (80.425
\pm 0.034)$ GeV \cite{exp}. Recently, a lot of progress has been made towards
establishing accurate theoretical prediction for $\MW$. The best result
\cite{mw} includes  the complete two-loop corrections \cite{muon,AC2} and
some three-loop contributions \cite{qcd3,mt6}. The remaining theoretical error
is estimated to be $\delta\MW \sim 4$ MeV, which is well below the current experimental
uncertainty. Still, the electroweak two-loop corrections total to
$\sim 30$ MeV and are thus mandatory for electroweak precision analyses.

The effective weak mixing angle \SinEff\ is mainly 
derived from various asymmetries measured around the $Z$ boson peak at
$e^+ e^-$ colliders after subtraction of QED effects.
The current experimental accuracy, 
$\sineff = 0.23150 \pm 0.00016$ \cite{exp} implies strong indirect constraints
on the allowed range for the Higgs boson mass $\MH$. Therefore it is important
to develop precise theoretical calculations for this quantity.

Usually, \SinEff\ is computed as a function of the electromagnetic coupling
$\alpha$, the muon constant $G_\mu$ and the masses of the $Z$ boson, $\MZ$, and
the top quark, $\mt$ (other fermion masses are numerically irrelevant). As
explained before, $\MW$ is calculated from $G_\mu$, but in addition to these
corrections, the computation of \SinEff\ involves the corrections to the $Z$
vertex form factors.
The latter are expressed by the quantity $\kappa = 1+\Delta\kappa$,
in such a way that the effective weak mixing angle can also be written as:
\begin{equation}
  \sineff = \left(1- M_W^2/M_Z^2 \right)  \left(1+
  \Delta\kappa \right),
  \label{eq:kappa}
\end{equation}
and at tree-level,  $\Delta\kappa = 0$. Higher-order corrections to \SinEff\ have
been under extensive theoretical study over the last two decades. Besides the
one-loop result~\cite{sirlin,1loop}, two- and three-loop QCD corrections are
available \cite{qcd2,qcd3,qcd3light}, but for the electroweak two-loop
contributions only partial results were known. By means of a large mass
expansion in the heavy top quark mass, the formally leading ${\cal
O}(\alpha^2 \mt^4)$~\cite{ewmtmh,ewmt4} and next-to-leading ${\cal O}(\alpha^2
\mt^2 \MZ^2)$~\cite{ewmt2} terms were computed. A part of the missing two-loop
contributions was incorporated by the complete electroweak two-loop corrections to
$\MW$ \cite{muon,AC2}. While these corrections effected a shift in $\MW$
of 4 MeV compared to the previously known ${\cal O}(\alpha^2 \mt^2 \MZ^2)$
contributions, the induced shift in \SinEff\ was very sizable, $\delta\sineff =
8 \times 10^{-5}$, thus implying that the missing two-loop terms in the form
factor $\Delta\kappa$ can be of similar order.

\section{ELECTROWEAK TWO-LOOP CORRECTIONS TO \boldmath \SinEff}

\begin{figure}[tb]
\begin{center}
\psfig{figure=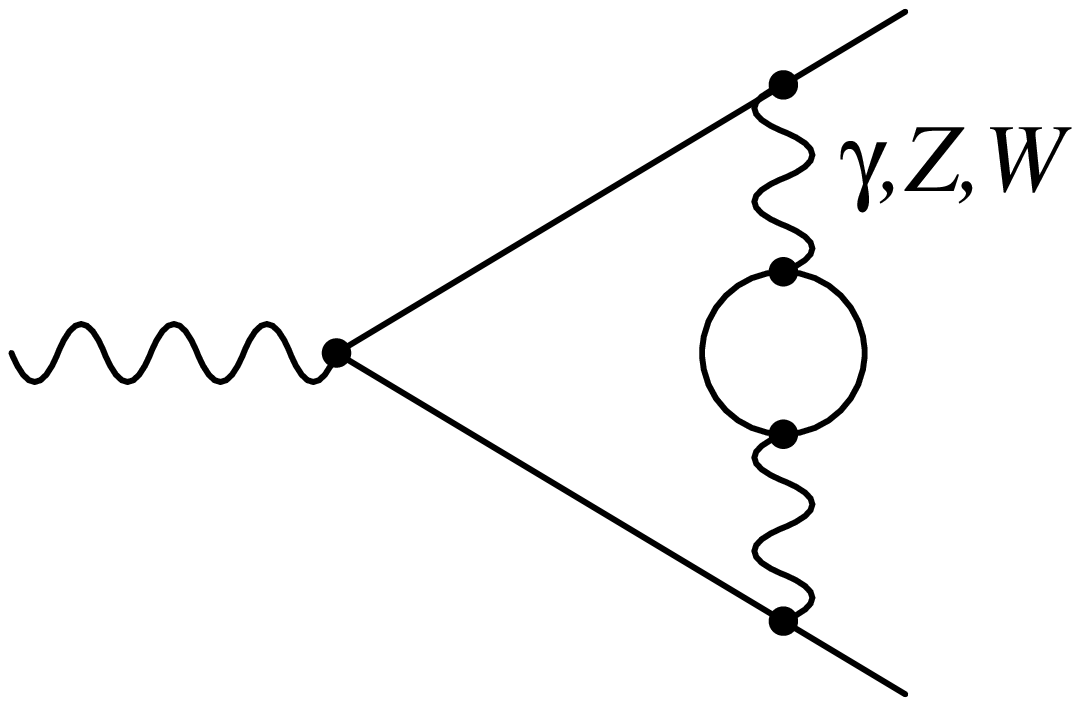, width=4cm} 
\psfig{figure=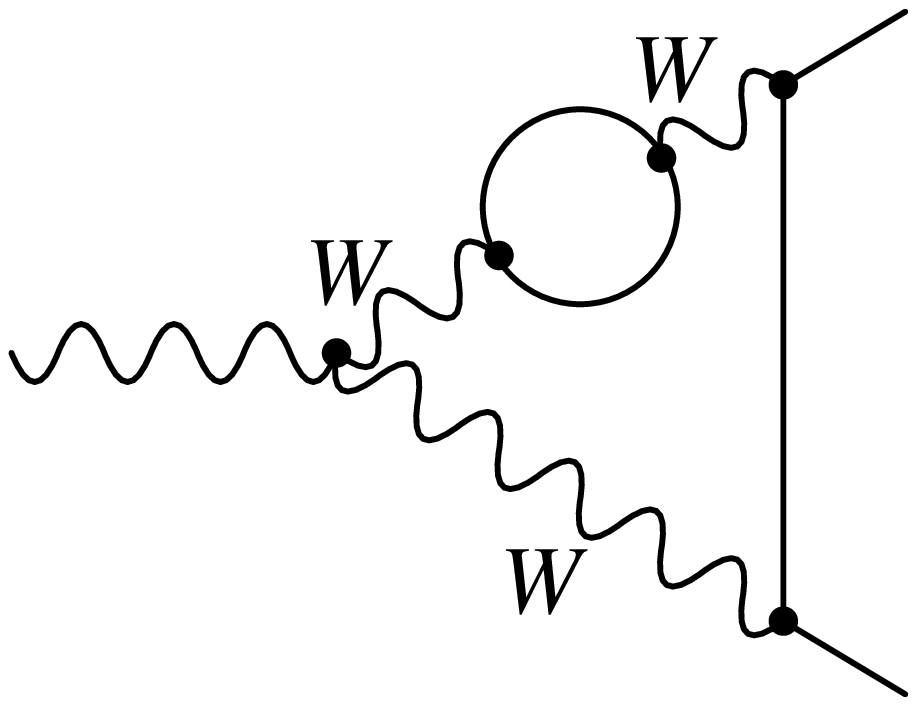, width=3.1cm} \hspace{1ex}
\psfig{figure=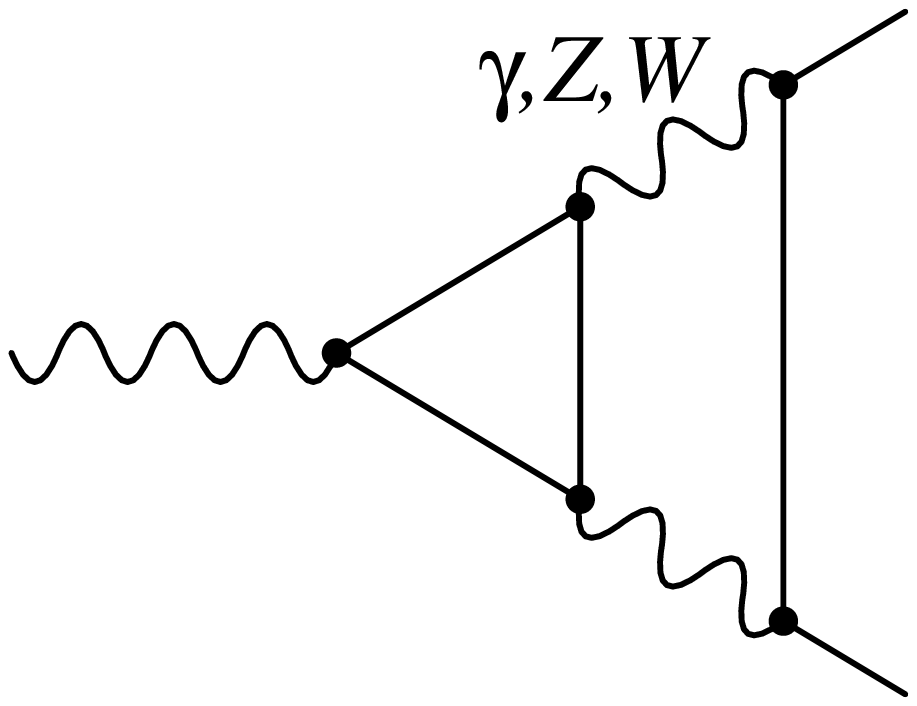, width=3.1cm} \hspace{1ex}
\psfig{figure=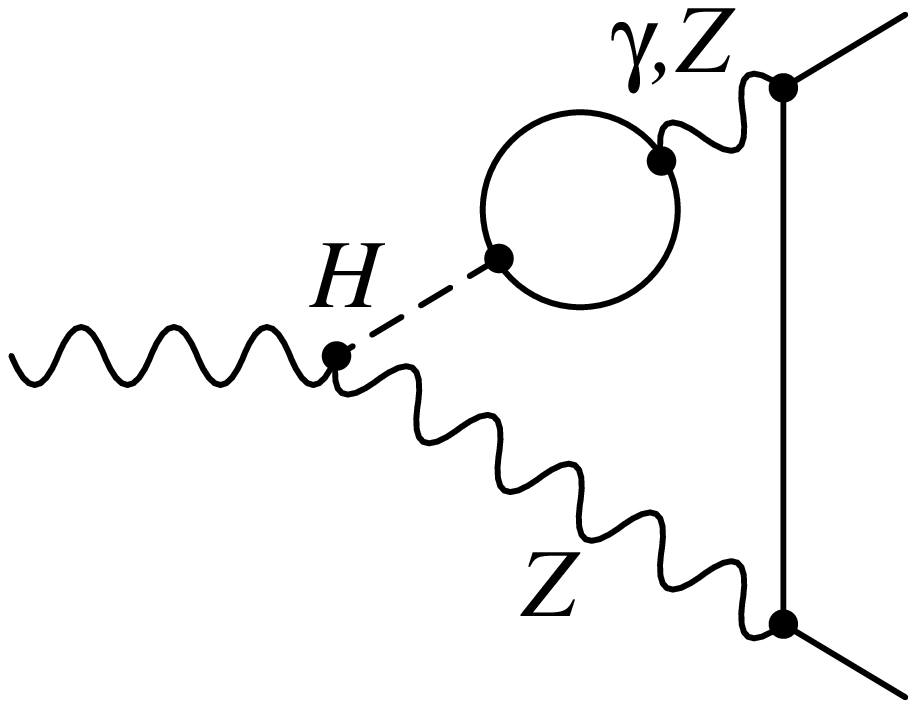, width=3.1cm} \hspace{1ex}
\end{center}
\vspace{-4ex}
\caption{
Genuine two-loop $Zl^+l^-$ vertex diagrams contributing to \SinEff.
\label{fig:diags}}
\end{figure}

As a first step towards completing the electroweak two-loop corrections to
\SinEff, results for the fermionic (i.e. diagrams with closed fermion loops)
two-loop corrections were presented recently~\cite{sineff}. The genuine
two-loop vertex  diagrams are represented by the generic topologies in
Fig.~\ref{fig:diags}.
Higher-order corrections to the process $e^+e^- \to f \bar{f}$ near the $Z$ pole
can be consistently computed by performing an expansion of the amplitude around
the complex pole ${\cal M}_{\rm Z}^2 = \MZ^2 - i \MZ \Gamma_{\rm Z}$,
\begin{equation}
{\cal A}[e^+e^- \to f \bar{f}] = \frac{R}{s-{\cal M}_{\rm Z}^2} + S + 
	(s-{\cal M}_{\rm Z}^2) S' +
\dots \label{eq:polexp}
\end{equation}
Here
$\Gamma_{\rm Z}$ is the $Z$ decay width. After subtracting
contributions from s-channel photon exchange and $\gamma$-$Z$ interference, the
vertex corrections form factor at NNLO is derived to be
\begin{equation}
\kappa^{(2)}_\Pf = 
\frac{\hat{a}_\Pf^{(2)} \, v_\Pf^{(0)} \, a_\Pf^{(0)} - 
   	\hat{v}_\Pf^{(2)} \, (a_\Pf^{(0)})^2 -
	(\hat{a}_\Pf^{(1)})^2 \, v_\Pf^{(0)} +
	\hat{a}_\Pf^{(1)} \, \hat{v}_\Pf^{(1)} \, a_\Pf^{(0)}}{(a_\Pf^{(0)})^2
	(a_\Pf^{(0)}-v_\Pf^{(0)})} \Biggr|_{s = \MZ^2},
\end{equation}
where the superscripts in parentheses indicate the loop order. In this quantity,
IR-divergencies from QED contributions drop out, which involves a delicate
interplay between one- and two-loop terms in the form factors.
The UV-divergencies are cancelled by on-shell renormalization. The relevant
counterterms are derived using the methods of Ref.~\cite{muon}.

The new part of this work is the computation of the two-loop $Zf\bar{f}$ vertex
corrections, which are treated with two independent technical methods. The first
method uses large mass expansions for the diagrams with internal top-quark lines
and the differential equation method for diagrams with only light fermions $f
\neq t$, the
masses of which are neglected. Contrary to previous work \cite{ewmt2}, the
expansion in $x = \MZ^2/\mt^2$ is performed to high precision, by executing the
series to the order $x^{10}$, reaching an overall relative precision of $\sim
10^{-5}$ of the final result. The coefficients of the expansion are 2-loop
tadpole and 1-loop vertex diagrams, which can be evaluated efficiently with
well-known analytical formulae. The contributions from diagrams without top-quark
propagators involve only two independent scales, $\MZ$ and $\MW$, allowing a
fully analytical treatment. Even for the limited set of diagrams with closed
fermions loops, a large number of scalar integrals with non-trivial structures
in the numerator are involved. They can be reduced to a set of scalar master
integrals by using integration-by-parts identities \cite{ibp}. 
Owing to size of the linear equation system associated with this reduction
procedure, the algorithm has been implemented in the dedicated C++
library \textsc{DiaGen/IdSolver} \cite{diagen}, which performs the necessary
steps in a highly automized way.
Analytical
results for these master integrals are obtained by the differential equation
method \cite{diffeq}. This is illustrated by the following example,
\begin{equation}
p^2 \frac{\rm d}{{\rm d}p^2} \left[
  \raisebox{-5.4mm}{\psfig{figure=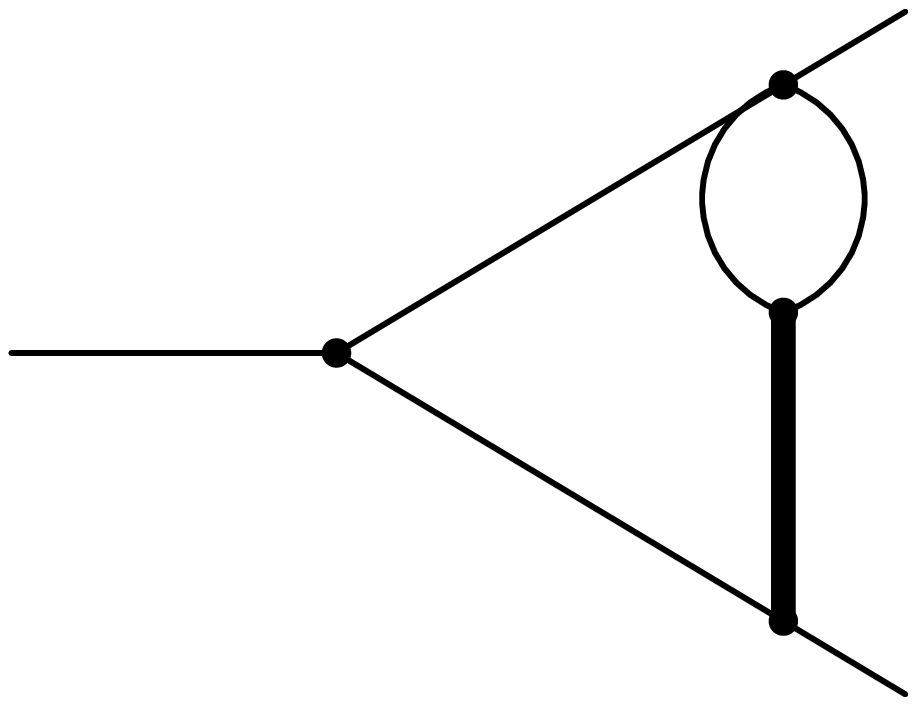, width=1.6cm}}\;\right]
 = 
  \frac{p^2}{p^2+m^2} \Biggl( \frac{4-D}{2}(4 + 5 \frac{m^2}{p^2})
  \left[ \raisebox{-5.4mm}{\psfig{figure=v2p1.ps, width=1.6cm}}\;\right]
+ \frac{10 - 3D}{2} \left[ \raisebox{-5.4mm}{\psfig{figure=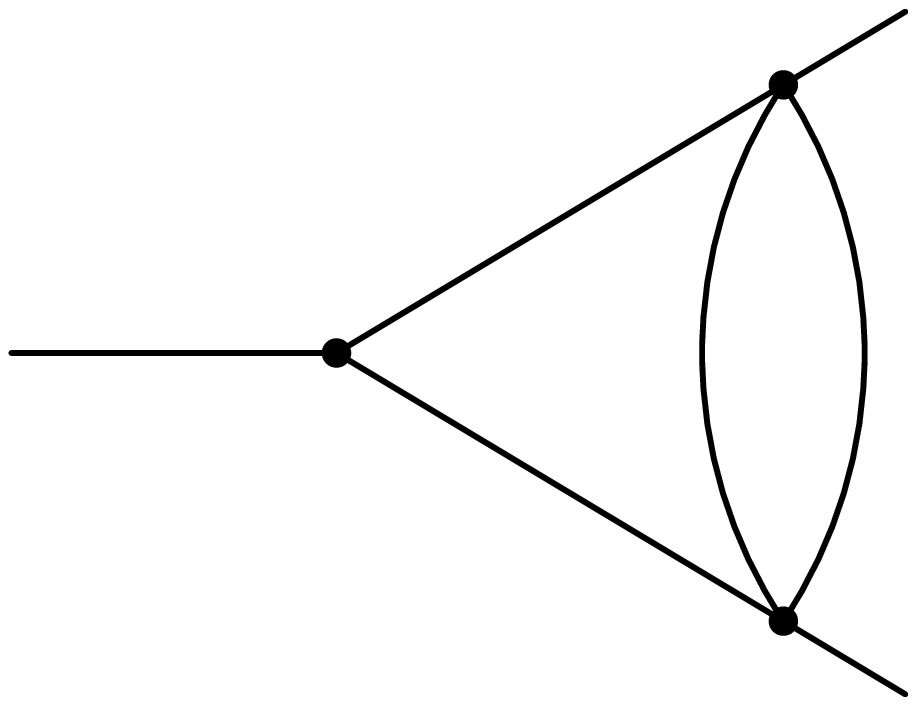, width=1.6cm}}\;\right]
  - \frac{2-D}{2} \left[ \raisebox{-3mm}{\psfig{figure=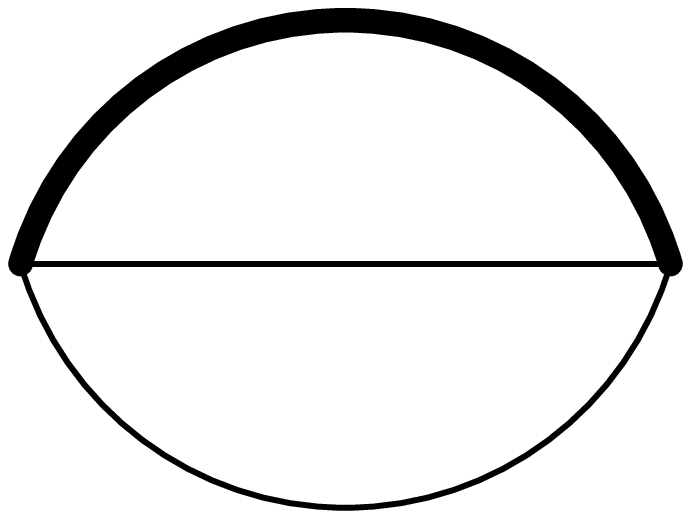, width=0.95cm}}\right]
  \Biggr).
\end{equation}
Here the thick lines represent massive propagators with mass $m$, the thin
lines denote massless propagators and $p$ is the momentum flowing into the
vertex. $D$ is the dimension of dimensional regularization.
The momentum derivate of the scalar integrals on the left-hand side
results in the same integral and simpler integral topologies on the right-hand
side. Feeding in analytical results for these simpler integrals, the
differential equation can be solved in terms of generalized polylogarithms.
All integrals were also checked by using low-momentum expansions.

\begin{figure}
\centering
\rule{0mm}{0mm}\vspace{-1.2ex}
\begin{tabular}{c@{\hspace{2cm}}c}
\epsfig{figure=sbau_pages.epsi, width=3.7cm, bb=210 450 330 530, clip=true} &
\psfig{figure=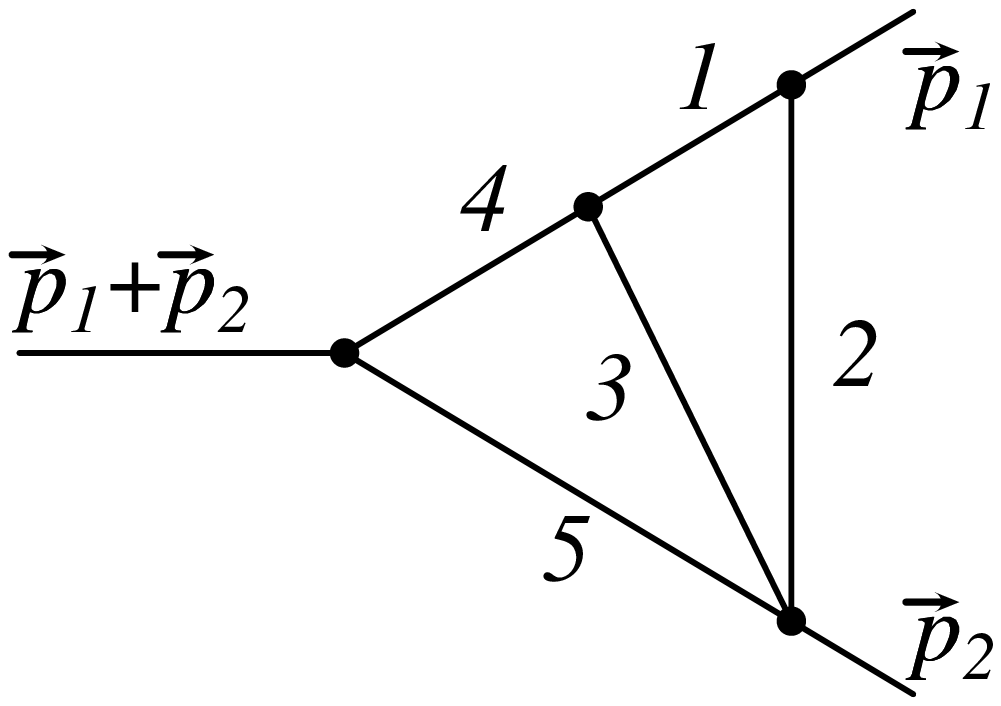, width=3cm}\raisebox{1.15cm}{$\!\!\!\to$} \
\psfig{figure=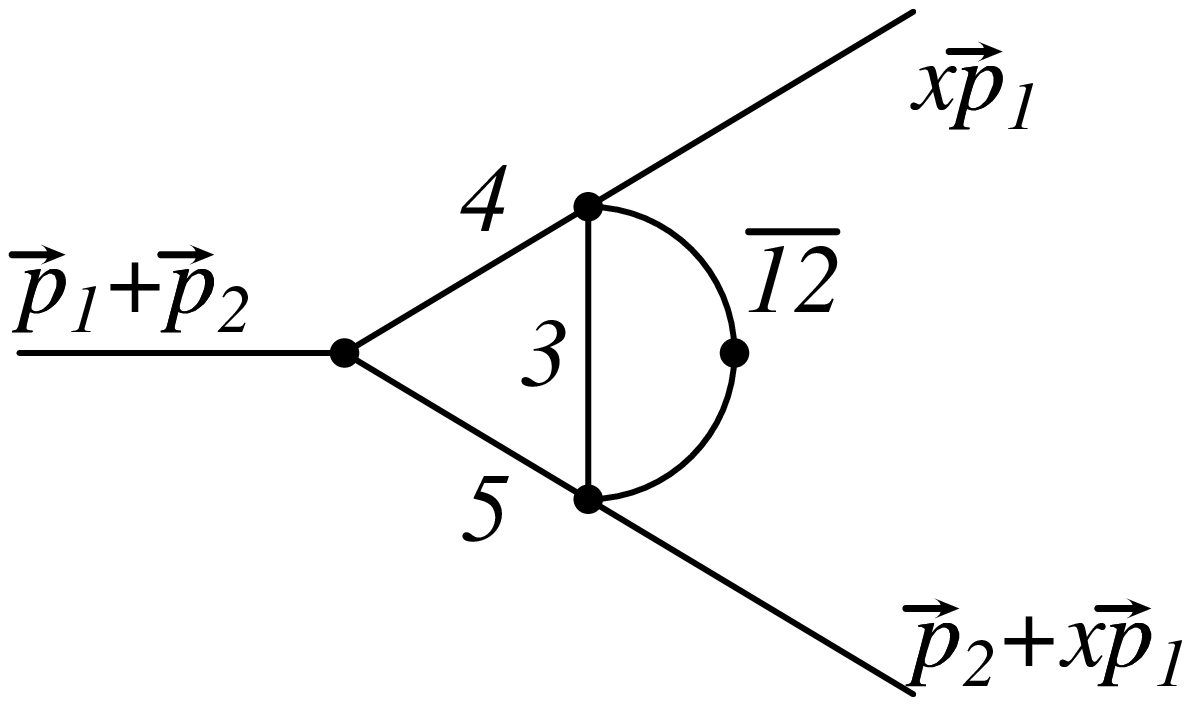, width=4cm} \\[-.5ex]
 (a) & (b) \\
\end{tabular}
\caption{(a) General representation of a two-loop scalar diagram with
self-energy sub-loop. (b) Reduction of triangle sub-loop to self-energy sub-loop
by means of Feynman parameters.}
\label{fig:disp}
\end{figure}

The second method makes use of numerical integrations based on dispersion
relations. A scalar two-loop integral with a self-energy sub-loop as in
Fig.~\ref{fig:disp}~(a) can be expressed as \cite{intnum}
\begin{equation}
T_{N+1}(p_i;m_i^2) = - \int_{s_0}^\infty {\rm d}s \;
  \Delta B_0(s,m_N^2,m_{N+1}^2) 
  \int {\rm d}^4 q  \,
  \frac{1}{q^2-s} \,
  \frac{1}{(q+p_1)^2 - m_1^2} \cdots \frac{1}{(q+p_1+\dots+p_{N-1})^2 -
  m_{N-1}^2},
\end{equation}
where $\Delta B_0$ is the discontinuity of the scalar one-loop self-energy
function. The second integral can be evaluated into a standard $N$-point
one-loop function, leaving the integration over $s$ to be performed
numerically. In general, one can also introduce dispersion relations for
triangle sub-loops \cite{intnum2}, but it is often  technically easier to
reduce them to self-energy sub-loops by introducing Feynman parameters
\cite{feynpar},
\begin{equation}
\begin{array}{c}
\ [(q+p_1)^2-m_1^2]^{-1} \; [(q+p_2)^2-m_2^2]^{-1} = \int_0^1 {\rm d}x
  \; [(q+\bar{p})^2 - \overline{m}^2]^{-2} \\
\bar{p} = x\,p_1 + (1-x)p_2, \qquad
\overline{m}^2 = x \, m_1^2 + (1-x) m_2^2 - x(1-x)(p_1-p_2)^2.
\end{array}
\end{equation}
This is indicated diagrammatically in Fig.~\ref{fig:disp}~(b). The integration
over the Feynman parameters is also performed numerically. As a result, all
master integrals for the vertex topologies can be evaluated by at most 3-dim.
numerical integrations. Before performing the numerical integrations, possible
UV- and IR-divergencies need to be subtracted from the integrals. While this
second method is applicable to two-loop vertex corrections with an arbitrary
number of mass scales, it is slower and leads to much large expressions than the
first method. Nevertheless it provides an important check of the result.

Special caution is needed for the diagrams with a fermion triangle loop (see
the third diagram in Fig.~\ref{fig:diags}), which involve the $\gamma_5$
matrix. In dimensional regularization, it is not possible to fulfill the two
relations $\{\gamma_\mu,\gamma_5\} = 0$ and
Tr$(\gamma^\alpha\gamma^\beta\gamma^\gamma\gamma^\delta\gamma_5) = 4 i
\epsilon^{\alpha\beta\gamma\delta}$ simultaneously. As in Ref.~\cite{muon}, 
the contributions resulting in $\epsilon$-tensors were therefore evaluated in
four dimensions, based on the observation that these terms are free of
UV-divergencies.  Potential soft and collinear divergencies of single diagrams
are regulated using a photon mass, with a subsequent careful expansion for zero
photon masses.

\section{RESULTS}

The new result for the fermionic two-loop corrections is combined with
corrections of order ${\cal O}(\alpha)$, ${\cal O}(\al\als)$ \cite{qcd2},
${\cal O}(\al\als^2)$ \cite{qcd3} and leading three-loop terms of ${\cal
O}(\al\als^2\mt^4)$ and ${\cal O}(\al^3\mt^6)$ \cite{mt6}. Reducible terms of
the same order are taken into account, but no resummations are preformed. The
most precise prediction of \SinEff\ is obtained as a function of the muon decay
constant $G_\mu$, from which the $W$-boson mass is calculated by including the
radiative corrections to $\MW$ as given in Ref.~\cite{mw}.

\begin{figure}[tb]
\begin{center}
\epsfig{figure=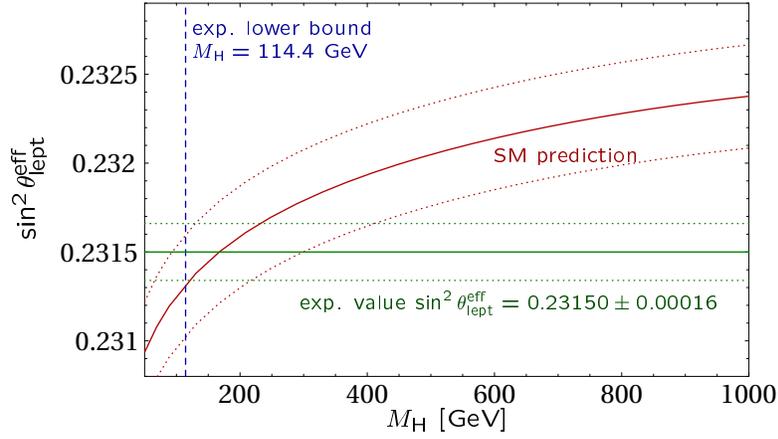, width=10cm,
        bb=40 218 651 571}
\end{center}
\vspace{-5ex}
\caption{Prediction for \SinEff\ including two-loop corrections compared to the
current direct experimental measurement, with 1$\sigma$ bands from experimental
input. The chosen input parameters are $\MZ = (91.1876 \pm 0.0021)$ GeV,
$\Gamma_{\rm Z} = 2.4952$ GeV, $\mt = 178.0 \pm 4.3$ GeV, $m_{\rm b} = 4.85$
GeV, $\Delta\alpha(\MZ^2) = 0.05907 \pm 0.00036$, $\als(\MZ^2) = 0.117 \pm
0.002$, $G_\mu = 1.16637 \times 10^{-5} \gev^{-2}$.}
\label{fig:res}
\end{figure}

Fig.~\ref{fig:res} shows the final result for \SinEff\ as a function of the Higgs mass
compared to the current experimental value.
Included in the plot are the error bands due to the uncertainties of the
experimental input parameters entering into the theoretical prediction and of
the direct measurement of \SinEff.  As evident from the figure, the
identification of computation and measurement for \SinEff\ favor relatively
small values for $\MH$. With the new two-loop corrections, the best-fit value
for $\MH$ moved from 148 GeV (using formulae of Ref.~\cite{ewmt2})
to 168 GeV (for $\mt = 178$ GeV).

The numerical result for the leptonic effective weak mixing angle has been
published in Ref.~\cite{sineff} as a parametric fitting formula that is
accurate in the range 10 GeV $\leq \MH \leq$ 1 TeV. It has also been
implemented into the newest version 6.42 of the program \textsc{Zfitter}
\cite{zfitter}, with some changes recently discussed in
Ref.~\cite{Freitas:2004mn}, and was used in the latest release of electroweak
precision global fits of the Standard Model. The impact of the new result for
\SinEff\ shifts the 95\% confidence level upper bound on the Higgs mass upwards
by 23 GeV to 260 GeV \cite{exp}.

Together with the inclusion of the new two-loop result in the prediction for
\SinEff, a assessment of the uncertainties from missing higher order
contributions is required. Since for practical purposes \SinEff\ is given as a
function of the muon decay constant $G_\mu$, it is useful to evaluate the
theoretical error for this parametrization, i.e. combining the radiative
corrections to $\MW$ and the $Z$ vertex.
A simple method to estimate the higher order uncertainties assumes 
that the perturbation series follows roughly a geometric progression.
This presumption implies relations like
$
{\cal O}(\al^2\als) = {\cal O}(\al^2)/{\cal O}(\al) \, {\cal O}(\al\als).
$
With this method one obtains the following errors for $\MH$ between 10 and 1000
GeV in units of $10^{-5}$: between 2.3 and 2.0 for the ${\cal O}(\alpha^2 \als)$
contributions beyond the leading $\mt^4$ term, between 1.8 and 2.5 for ${\cal
O}(\alpha^3)$, between 1.1 and 1.0 for ${\cal O}(\alpha \als^3)$ and between 1.7
and 2.4 for ${\cal O}(\alpha^2 \als^2)$. The missing bosonic ${\cal
O}(\alpha^2)$ corrections cannot be appraised from geometric progression.
However, considering they have a prefactor $\al^2$ but no specific enhancement
factor, they are estimated to be about $1.2 \times 10^{-5}$.
To account for possible deviations from the geometric series behavior, an
overall factor $\sqrt{2}$ was included to arrive at a total error of 
$\delta_{\rm th}\!\sineff = 4.9 \times 10^{-5}$.

Alternatively, the error from a higher-order QCD loop can be assessed by
varying the scale of the strong coupling constant $\als$ or the top-quark mass
$\mt$ in the $\overline{\rm MS}$ scheme in the highest available perturbation
order. The scale variation leads to an error estimate of 0.1 to $3.9 \times
10^{-5}$ for the ${\cal O}(\al^2\als)$ corrections and of less than $10^{-6}$
for the ${\cal O}(\al\als^3)$ contributions. These numbers are of the same
order as the estimated errors from the geometric progression method, so that the
total error given above seems to be fairly reliable.

The new error estimate was used in the latest electroweak global fits \cite{exp}
and lead to a reduction of the width of the well-known {\it blue band}, which
indicates the theoretical error in the indirect determination of the Higgs mass.

\section{CONCLUSIONS}

In this contribution, recent progress in the calculation of higher-order
corrections to the most important electroweak precision observables and their
impact on the indirect determination of the Higgs mass was reported.

The complete fermionic ${\cal O}(\alpha^2)$ corrections to the leptonic effective
weak mixing angle \SinEff\ have been calculated and numerical results were
presented. As an additional check,
the computation of the two-loop vertex integrals was performed with
two independent methods. The new result, together with an estimate of the
remaining theoretical error, was included in the latest version of the program
\textsc{Zfitter} and used for the latest global electroweak fits of the Standard
Model. 

The calculation of the remaining bosonic electroweak two-loop corrections is
currently in progress and will be available soon \cite{future}. Furthermore, we
are working on adapting the new results for quark final states, where particular
attention has to be paid to the $Zb\bar{b}$ vertex, since it includes additional
massive top-quark propagators \cite{future}.

\begin{acknowledgments}
The authors are grateful to G.~Weiglein for valuable contributions, comments and
communications concerning this project.
M.~C.  was supported by the Sofja Kovalevskaja Award of the Alexander
von Humboldt
Foundation sponsored by the German Federal Ministry of Education and
Research.  M.~A. and M.~C. were supported by
the Polish State Committee for Scientific Research (KBN)
for the research project in years 2004-2005.
\end{acknowledgments}


\end{document}